\documentclass[reprint]{agujournal2019}
\usepackage{url} 
\usepackage{lineno}
\usepackage{soul}
\usepackage{amssymb, amsmath, amsthm, amsfonts}
\usepackage{multirow} 
\usepackage{graphicx} 

\draftfalse

\journalname{Geophysical Research Letters}

\begin{document}

\title{Impact of climate change on surface stirring and transport in the Mediterranean Sea}

%
%




\authors{Enrico Ser-Giacomi * \affil{1,2}  }

\authors{Gabriel Jord\'{a} S\'{a}nchez \affil{3}}
\authors{Javier Soto-Navarro \affil{4} }
\authors{S\"{o}ren Thomsen \affil{2} }
\authors{Juliette Mignot \affil{2} }
\authors{Florence Sevault \affil{5}}
\authors{Vincent Rossi \affil{6}}
\authors{$\quad$}

\affiliation{1}{Department of Earth, Atmospheric and Planetary Sciences, Massachusetts Institute of Technology, 54-1514 MIT, Cambridge, MA 02139, USA.}
\affiliation{2}{Sorbonne Universites, LOCEAN-IPSL, IRD, CNRS, UMR 7159, Paris, France}
\affiliation{3}{Instituto Espa\~nol de Oceanograf\'{i}a, Centre Oceanogr\'{a}fic de Balears, Moll de Ponent 07015 Palma, Spain}
\affiliation{4}{Mediterranean Institute for Advances Studies (IMEDEA, UIB-CSIC), Mallorca (Spain)}
\affiliation{5}{CNRM, Toulouse University, M\'{e}t\'{e}o-France, CNRS, Toulouse, France}
\affiliation{6}{Mediterranean Institute of Oceanography (UM 110, UMR 7294), CNRS, Aix Marseille Univ., Univ. Toulon, IRD, 13288, Marseille, France.}

\correspondingauthor{Enrico Ser-Giacomi}{enrico.sergiacomi(at)gmail.com}




\begin{keypoints}

\item We exploit a coupled climate model over the Mediterranean combining Network Theory with a Lagrangian approach.

\item Entropy and kinetic energies analyses project a significant increase of stirring for the next century.

\item Future transport patterns result in larger areas variability and stronger internal mixing of hydrodynamic provinces.

\end{keypoints}


\begin{abstract}

Understanding how climate change will affect oceanic fluid transport is crucial for environmental applications and human activities. However, a synoptic characterization of the influence of climate change on mesoscale stirring and transport in the surface ocean is missing. To bridge this gap, we exploit a high-resolution, fully-coupled climate model of the Mediterranean basin using a Network Theory approach. We project significant increases of horizontal stirring and kinetic energies in the next century, likely due to increments of available potential energy. The future evolution of basin-scale transport patterns hint at a rearrangement of the main hydrodynamic provinces, defined as regions of the surface ocean that are well-mixed internally but with minimal cross-flow across their boundaries. This results in increased heterogeneity of province sizes and stronger mixing in their interiors. Our approach can be readily applied to other oceanic regions, providing information for the present and future marine spatial planning.

\end{abstract}


\section*{Plain Language Summary}

Transport and mixing of water masses driven by ocean currents influences a variety of fundamental processes, including heat redistribution, ecosystem functioning and pollutants spreading. Therefore, understanding how fluid transport will be affected by climate change is crucial, in particular in the ocean surface, where marine life as well as human activities are concentrated. Here, we exploit a state-of-the-art climate model over the Mediterranean basin using a novel methodology which integrates Network Theory concepts with Lagrangian modeling. We assess past conditions and future changes at climatic scales of ocean stirring and transport over the entire basin. Our results reveal a significant increment of surface stirring linked to an increase of currents kinetic energy, which in turn could be ascribed to increments of available potential energy. We then provide a regionalization of the ocean surface based on hydrodynamic provinces that are well-mixed internally but with little leaking across their boundaries. Our model project an increased heterogeneity of province sizes and a stronger mixing in their interiors, while their mean area and coherence remain unaffected. Our approach could be applied to other oceanic domains and help designing adaptive strategies for marine spatial planning.




\section{Introduction}

In the oceanic environment, the transport and mixing of water masses driven by three-dimensional multiscale ocean currents play fundamental roles such as regulating the global climate and structuring marine ecosystems. A major contributor of transport and mixing is the so-called process of ``horizontal stirring". It is mostly driven (except at very small or large scales) by vigorous mesoscale currents \cite{ferrari2009ocean} (typical scales range 10-100 km and 10-100 days) which bear a large part of the horizontal kinetic energy \cite{chelton1998geographical,corrado2017general}. Importantly, mesoscale horizontal stirring, along with its spatio-temporal variability, is most relevant to study the dynamics of marine ecosystems \cite{rossi2008comparative, mcgillicuddy2016mechanisms,dubois2016linking}, the evolution of the regional climate \cite{gutowski2016wcrp}, the tracers redistribution and the fate of pollutants \cite{zhang2014oceanic} as well as the transfer efficiency of heat and gas across the oceanic boundary layer \cite{frenger2013imprint}. Hence, we here focus on the surface ocean not only for its transitional and highly-energetic characters, but also because it is where most marine life and human activities occur. 

While transport and mixing have been extensively studied under the present climate, a systematic characterization of their future evolution with climate change is lacking. This is mainly due to the limitation of coarse climate models that do not resolve mesoscale eddies. Indeed, although the typical spatial resolution of the ocean module of most coupled climate models has been increasing from $\sim 1-2^\circ$ to $\sim 0.5^\circ$, this is still not enough to properly resolve mesoscale dynamics. Moreover, classical Lagrangian approaches \cite{ott2002chaos,d2004mixing,beron2008oceanic,hernandez2012seasonal}, that represent the ``golden standard'' to explicitly characterize transport phenomena in geophysical flows, do not provide  a comprehensive description of stirring and transport as they are rather dedicated to capturing ephemeral frontal or vortical structures focusing on specific scales. Furthermore, while recent research aims at developing alternative methodologies to better comprehend the complexity and multi-scale character of fluid transport processes, such as set-oriented \cite{froyland2003detecting,froyland2014well,miron2017lagrangian} or flow networks approaches \cite{rossi2014hydrodynamic,ser2015flow,lindner2017spatio,padberg2017network,wichmann2019mixing}, they have not been applied yet to long-term climate model outputs.

Assessing the future evolution of surface transport and mixing gains a particular relevance in relatively small and densely populated marginal seas with complex bathymetry such as the Mediterranean Sea. In fact, the Mediterranean Sea displays all the typical features of the oceanic circulation  \cite{millot2005circulation} and is characterized by an exceptional and threatened biodiversity \cite{coll2010biodiversity}. Moreover, it is a well-recognized climatic hot-spot being one of the most responsive region to climate change on Earth \cite{giorgi2006climate,diffenbaugh2012climate}. For such reasons the Mediterranean basin represents an ideal benchmark to investigate climate change effects on horizontal  stirring and transport and to provide an approach that can be  generalized to other marginal seas or oceanic region such as the Southern Ocean \cite{naveira2011eddy} or the eastern-boundary upwelling systems \cite{rossi2008comparative,ndoye2017dynamics} and possibly to other geophysical fluids \cite{ploeger2015quantifying}.

To contribute to bridging this gap, we exploit the outputs of a regional climate model run for the Mediterranean basin  by the Med-CORDEX initiative \cite{ruti2016med}. The high resolution of this configuration and the full coupling among its ocean, atmosphere and land components provide an unprecedented opportunity to properly address possible changes in transport horizontal patterns at climatic scales \cite{oerder2016mesoscale,frenger2013imprint}. To this aim, we use the Lagrangian Flow Network (LFN) approach \cite{ser2015flow,donner2019characterizing}, which is based on a probabilistic description of fluid transport \cite{ser2015most, ser2015dominant}. It suits well with the need to extract reliable long-term signals at different spatio-temporal scales \cite{rossi2014hydrodynamic}, hence characterizing transport phenomena in various, yet complementary, manners. Indeed, for each simulation, typical advection durations can be chosen to match any temporal scale of interest. Spatial scales can be explored both at (i) the \emph{local} scale associated with the size of single nodes of the network and at (ii) the \emph{global} scale of the main network communities.  This methodology has solid theoretical foundations and has been already compared with alternative approaches \cite{ser2015flow}. Moreover, it has been tested for its sensitivity and robustness \cite{monroy2017sensitivity} and successfully used in disparate applications \cite{dubois2016linking,hidalgo2019accounting,legrand2019multidisciplinary}.

We concentrate here on two disconnected time windows spanning several decades; one represents the present climate and another one stands for the future climate projected under the Representative Concentration Pathway (RCP) 8.5 scenario, in accord with the IPCC Special Report on Ocean and Cryosphere \cite{portner2019ipcc}. In each temporal window, thousands of basin-wide Lagrangian simulations provide a robust description of transport processes, focusing on the typical time scales of mesoscale variability \cite{beron2008oceanic,hernandez2012seasonal}. At local spatial scales, we address patterns and trends of horizontal stirring by calculating node entropies (Section \ref{sec:re_entropy}) and we unveil their relations with kinetic energy and available potential energy (Section \ref{sec:res_entrenerg}). While at global scale, we study how the horizontal mean and turbulent flows delineate synoptic hydrodynamical provinces i.e. an ensemble of oceanic regions in which water parcels are well-mixed and mostly confined. We finally study their evolution between the present and the future model runs and we discuss the possible causes and implications of such changes (Section \ref{sec:res_provinces}).


\section{Materials and Methods}
An extended description of the methodology, including mathematical formulations and technical details, is provided in the Supplementary Information, Section 1.

\subsection{Velocity fields data from Med-CORDEX}\label{subsec:medcordexdata}
Med-CORDEX is an initiative for coordinated high resolution Regional Climate Models (RCMs) simulations over the Mediterranean basin where the ocean model component is fully-coupled with the atmospheric one \cite{ruti2016med}. We use the daily velocity fields of the ocean component (NEMOMED8) of the CNRM-RCSM4 model at 9.8 m depth and at $1/12^{\circ}$ of horizontal resolution \cite{sevault2014fully}. This model aims at reproducing the regional climate system with as few constraints as possible. Two different runs are used here: the first simulates the historical climate (1950-2005) and the second projects the future climate under the high CO2 emission RCP8.5 forcing scenario (2005-2100). For comparative purposes we analyze time-periods of three decades for each run (1971-2000 and 2071-2100,  respectively). This model configuration has been already used in other studies and the historical simulation was intensively validated against observations prior to using the scenario period \cite{darmaraki2019future, soto2020evolution}.

\subsection{Lagrangian Flow Networks construction}\label{subsec:lfnsetup}
We build Lagrangian Flow Networks \cite{ser2015flow,ser2015most} (LFNs) using the Med-CORDEX near-surface velocity fields. The entire Mediterranean surface is subdivided into $N = 3591$ two-dimensional network nodes of linear size $\Delta = 27.78$ km. Each node is uniformly filled with around 600 Lagrangian particles (or less for coastal regions, proportionally to the land cover of each node). A network is uniquely characterized by the \emph{starting time} $t_0$ when particles are seeded and the \emph{integration time} $\tau$ of their Lagrangian trajectories. All transport and mixing features are encoded in the \emph{adjacency matrix} $\mathbf{P}(t_{0},\tau)$ associated with each network. The matrix element  $\mathbf{P}(t_{0},\tau)_{ij}$, representing the weight of the link joining node $i$ to node $j$, is proportional to the number of particles  (equivalent to water parcels) whose trajectory started from location $i$ at time $t_0$ and ended in node $j$ at time $t_0 + \tau$.

To highlight projected changes of stirring and transport patterns at climatic scales, we build LFNs using $t_0$ at a weekly frequency over the two temporal windows representative of the historical and future runs, respectively. This leads thus to 2880 different starting times $t_0$ and allows to consider intra-seasonal, seasonal and inter-annual patterns between the two runs with a sufficient statistical power. However, we focus hereafter on changes between the historical and scenario runs while the variability associated to shorter time scales is implicitly accounted by fluctuations around mean values of the analyzed measures. We consider integration times $\tau$ of 30, 60 and 90 days.

\subsection{Network entropies and kinetic energies}\label{subsec:entropyandkindef}
Network entropies have been introduced as a family of diagnostics explicitly based on Lagrangian trajectories that accurately quantify horizontal stirring at the scale of single nodes \cite{ser2015flow,lindner2017spatio}. They depend on a parameter $q$ that controls how much fluid volumes are taken into account in the stirring calculation. For our analysis we fix the parameter $q$ equal to one, obtaining a Shannon-like entropy that can be calculated on the incoming links (backward in time dynamics) or on the outgoing ones (forward in time dynamics) \cite{ser2015flow,ser2017lagrangian,wichmann2019mixing}. The expression for the in-entropy $H^{I}(t_0,\tau)_i$ and the out-entropy $H^{O}(t_0,\tau)_i$ are:
\begin{align}
  H^{I}(t_0,\tau)_i &= -\frac{1}{\tau} \sum_{j=1}^{N} \bigg( \frac{\mathbf{P}(t_{0},\tau)_{ji}}{\sum_{k=1}^{N} \mathbf{P}(t_{0},\tau)_{ki}} \bigg) \log \bigg( \frac{\mathbf{P}(t_{0},\tau)_{ji}}{\sum_{k=1}^{N} \mathbf{P}(t_{0},\tau)_{ki}}\bigg) ,\\
 H^{O}(t_0,\tau)_i &= -\frac{1}{\tau} \sum_{j=1}^{N} \bigg( \frac{\mathbf{P}(t_{0},\tau)_{ij}}{\sum_{k=1}^{N} \mathbf{P}(t_{0},\tau)_{ik}} \bigg) \log \bigg( \frac{\mathbf{P}(t_{0},\tau)_{ij}}{\sum_{k=1}^{N} \mathbf{P}(t_{0},\tau)_{ik}}\bigg) .
\end{align}
Therefore, the in-entropy is a weighted measure of the origins diversity of the water arriving at node $i$ at time $t_0+\tau$; the out-entropy measures instead the diversity of the destinations of the water present in the node $i$ at time $t_0$. We define (and use hereafter) the symmetrized-in-time entropy as an average of $H^{I}(t_0,\tau)$ and $H^{O}(t_0,\tau)$:
\begin{equation}
H(t_0,\tau)_i = \frac{1}{2} \Big( H^{I}(t_0,\tau)_i + H^{O}(t_0,\tau)_i \Big)
\end{equation}\label{eq:entropydef}
and we will simply call it \emph{Entropy}.

To compare entropy with widely used Eulerian diagnostics, we compute the \emph{Kinetic Energy} (KE) as $(u^2+v^2)/2$ at daily frequency. We then make temporal averages over $[t_0;\tau]$ intervals and we spatially average the KE field over each node of the network. We also consider the yearly \emph{Mean Kinetic Energy} (MKE) $\frac{<u>^2+<v>^2}{2}$ and the yearly \emph{Eddy Kinetic Energy} (EKE) $\frac{<u'^{\,2}>+<v'^{\,2}>}{2}$, with $u'=u-<u>$, $v'=v-<v>$ and $<>$ being a temporal mean over a year of daily velocities. MKE and EKE are also spatially averaged over each node of the network. KE, MKE and EKE are expressed in $m^2s^{-2}$. Finally, the \emph{Available Potential Energy} (APE), associated to the inverse of the Richardson number \cite{green1970transfer,stammer1998eddy}, is estimated as: $-\frac{f^2}{\rho_0} \frac{\big(\frac{\partial \rho}{\partial x}\big)^2 + \big(\frac{\partial \rho}{\partial y}\big)^2}{\frac{\partial \rho}{\partial z}}$.

\subsection{Hydrodynamic provinces}\label{subsec:provdef}
We identify \emph{hydrodynamic provinces} in the Mediterranean Sea as network communities in LFNs \cite{ser2015flow}. Such provinces are regions of the surface ocean which are well-mixed internally but with little leaking across its boundaries over specific period of time \cite{rossi2014hydrodynamic}. To provide optimal partitions of the Mediterranean basin in hydrodynamic provinces we use the Infomap algorithm \cite{rosvall2008maps}. Infomap demonstrated indeed to perform better to other approaches when, as the case studied here, several scales are interacting simultaneously and no information is available on the expected number of provinces \cite{ser2015flow}. For each given $t_0$ and $\tau$ we can define thus a unique partition of the entire basin in different provinces (Supplementary Fig. 8, panel a).

Building upon the earlier work by \cite{rossi2014hydrodynamic,ser2015flow}, we use metrics to evaluate the dynamical properties of our hydrodynamic provinces. For a province $A$ we compute its \emph{coherence ratio} $\rho_{t_0}^{\tau}(A)$ that is the fraction of particles that at time $t_0+\tau$ are found in the same province where they were released at initial time $t_0$ \cite{ser2015flow}. We also compute the \emph{mixing parameter} $\mu_{t_0}^{\tau}(A)$ that measures how strongly the flow mixes fluid inside a province $A$ \cite{ser2015flow}. Hence, while the coherence ratio quantifies how much a province is able to retain fluid particles inside its boundary, the mixing parameter evaluates instead how much a province is internally well connected (Supplementary Fig. 8, panels b and c).

To quantify explicitly the effective retention of provinces boundaries (that is correlated to the efficiency of the associated transport barrier), we introduce a new metric based on a symmetric probability of water segregation between each pair of provinces $A$ and $B$, called \emph{boundary strength} and defined as:
\begin{equation}\label{eq:bstrdef}
\sigma_{t_0}^{\tau}(A,B) = \bigg(1-\frac{\sum_{i \in A ; j \in B} \mathbf{P}(t_{0},\tau)_{ij}}{\sum_{i \in A} \mathbf{P}(t_{0},\tau)_{ij}}\bigg) \bigg(1-\frac{\sum_{i \in B ; j \in A} \mathbf{P}(t_{0},\tau)_{ij}}{\sum_{i \in B} \mathbf{P}(t_{0},\tau)_{ij}}\bigg) .
\end{equation}
Note that, by definition, $\sigma_{t_0}^{\tau}(A,B) \in [0,1]$ and $\sigma_{t_0}^{\tau}(A,B)$ will be equal to 1 only when no water exchange occurred among $A$ and $B$ in the interval $[t_0;\tau]$. In fact, if we randomly pick up a fluid particle in $A$ and another in $B$ at time $t_0$, the boundary strength corresponds exactly to the probability that, after a time $\tau$, none of the two particles crossed the boundary between $A$ and $B$. Thus, the boundary strength tells how much the boundary between province $A$ and $B$ is ``impermeable'' to fluid particles and how much it can prevent fluid exchanges among the two provinces (Supplementary Fig. 8, panel d).

\subsection{Statistics on multiple partitions}\label{subsec:provmultiplepart}

To aggregate information from several partitions (i.e. several values of $t_0$ and $\tau$) we introduce new metrics to describe the mean geometry and coherence of all boundaries of a given partition. Let's consider $M$ different partitions associated to a set of $[t_0^{\alpha};\tau^{\alpha}]$ intervals with $\alpha = \{1, ... , M\}$. We call $A_i^{\alpha}$ a province in the partition $\alpha$ which the node $i$ belongs to. 

Given $M$ partitions and $N$ nodes, we can define the \emph{global coherence ratio} across the entire Mediterranean Sea as the spatial and temporal mean of each $\rho (A^{\alpha}_i)$:
\begin{equation}
\bar{\langle \rho \rangle} =  \frac{1}{NM} \sum_{i}^{N} \sum_{\alpha}^{M} \rho (A^{\alpha}_i),
\end{equation}
and corresponds to the coherence of the entire basin averaged across the set of $M$ $\alpha$-partitions. Following a similar procedure, we can define the \emph{global mixing parameter} as the spatial and temporal mean of each $\mu (A^{\alpha}_i)$:
\begin{equation}
\bar{\langle \mu \rangle} =  \frac{1}{NM} \sum_{i}^{N} \sum_{\alpha}^{M} \mu (A^{\alpha}_i),
\end{equation}\label{eq:basinmixpar}
and the \emph{global boundary strength} as the spatial and temporal mean of each $\sigma (A^{\alpha}_i)$:
\begin{equation}
\bar{\langle \sigma \rangle} =  \frac{1}{NM} \sum_{i}^{N} \sum_{\alpha}^{M} \sigma (A^{\alpha}_i).
\end{equation}
Note that our multi-partition approach is more statistically reliable than others based on a single matrix for two main reasons: (i) it asses explicitly the intrinsic variability due to different initial conditions; (ii) the temporal overlap across simulations ensures the robustness of the analysis to the possible instability of the clustering solutions.

\subsection{Heuristic relations between province areas and perimeters}\label{subsec:area_per}
We finally study the variation of the standard deviation of areas and total perimeter of a set of 2-dimensional shapes (e.g. hydrodynamic provinces in the Mediterranean) when the distribution of areas evolve. We introduce a set of $N^a$  shapes with associated areas $\{a_i\}$ and shape factor $s$ (the factor linking perimeter with area). Their  total perimeter will be $p^a =  s\sum_{i=1}^{N^a} \sqrt{a_i}$ and the area standard deviation will be $\sigma^a = \sqrt{\frac{1}{N^a} \sum_{i=1}^{N^a} (a_i^2) - (\mu^a)^2}$. 	We transform the set of $\{a_i\}$ into a new set  $\{b_i\}$ operating multiple exchanges of area among pairs of shapes $i$-$j$ following the area-preserving rule: $b_i = a_i + \sum_{j=1}^{N^a}\epsilon_{ij}$. Keeping constant the number of shapes and their mean area implies that $\sigma^b > \sigma^a \Leftrightarrow \sum_{i=1}^{N} b_i^2 > \sum_{i=1}^{N} a_i^2$. Assuming small area changes, we find:
\begin{align}
&\sigma^b - \sigma^a \simeq  2 \sum_{i=1}^{N} \sum_{j=i}^{N} \epsilon_{ij} (a_i - a_j) \label{eq:per_sd_bshape1} \\
&p^{b} - p^{a} \simeq \frac{1}{2} \sum_{i=1}^{N} \sum_{j=i}^{N} \epsilon_{ij} \bigg(\frac{1}{\sqrt{a_i}} -\frac{1}{\sqrt{a_j}} \bigg)\label{eq:per_sd_bshape2}
\end{align}
Therefore, when large shapes become larger by eroding area from shapes smaller than them, we expect an increase of standard deviation and a decrease of total perimeter. Conversely, when small shapes become larger by eroding area from shapes larger than them, we expect instead a decrease of standard deviation and an increase of total perimeter.


\section{Results and Discussion}

\subsection{Local scale: patterns of entropy and kinetic energy increase}\label{sec:re_entropy}

The analysis of the entropy introduced in Section \ref{subsec:entropyandkindef} provides a robust description of horizontal stirring at local scale ($\sim 25$ km) and can be seen intuitively as a weighted measure of the diversity of destinations (forward-in-time) and origins (backward-in-time) of the water particles contained in a node \cite{ser2015flow,ser2017lagrangian}.

\begin{figure}[h]
    \centering\includegraphics[width=\textwidth]{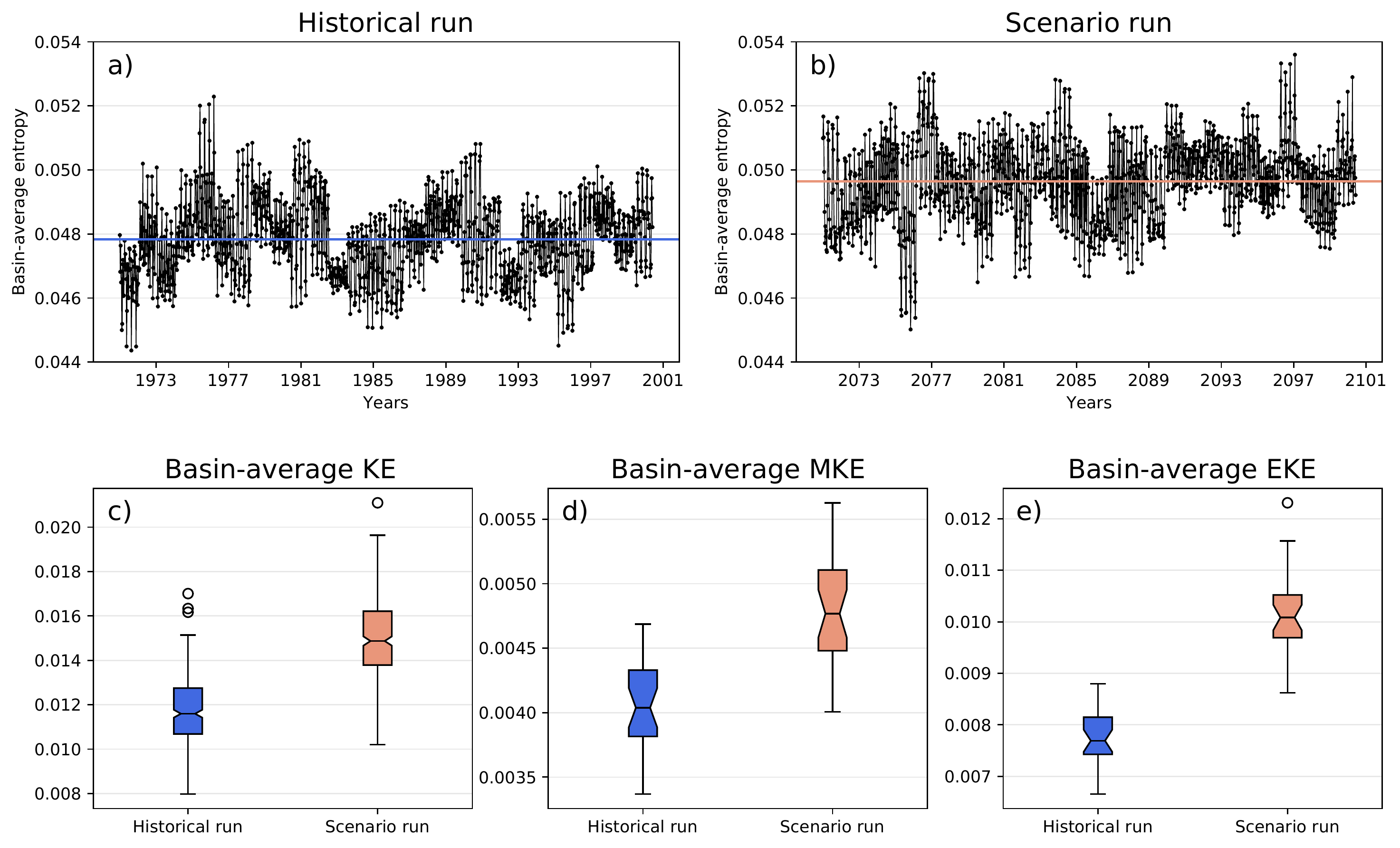}

\caption{\emph{Changes of basin-averaged entropy and kinetic energies}. Upper panels show time-series of the basin-scale spatial averages of entropy for the historical run a) and the future run b). Lower panels display, for both historical (blue) and scenario (orange) runs, the notched boxplots of the basin-scale averages of KE c), MKE d) and EKE e). Boxplot whiskers represent the 1.5 interquantile ranges. Notches represent the 95\% confidence interval around the median calculated from bootstrap.}  \label{fig:entropy_time_series60d}
  
\end{figure}

The weekly time-series of the basin-scale spatial average of entropy for $\tau=60$ days are shown in Fig. \ref{fig:entropy_time_series60d}, panels a) and b) (Section \ref{subsec:lfnsetup}). When computing the temporal mean of such basin averaged entropy over both temporal windows, we find an extremely significant (p-value $<10^{-4}$) increase of stirring of about 3.7\% in the scenario run with respect to the historical run. Conversely, the associated standard deviations in both runs are comparable. Similar mean changes are consistently found for other integration times:  3.7\% for $\tau=30$ days and 3.3\% for $\tau=90$ days. These results are backed-up by the statistical distribution of kinetic energy (KE), mean kinetic energy (MKE) and eddy kinetic energy (EKE) (Section \ref{subsec:entropyandkindef}) across both runs. We find indeed extremely significant increases of the temporal mean of their basin-averaged values : 27\% for KE, 17\% for MKE and 33\% for EKE (Fig. \ref{fig:entropy_time_series60d}, panels c), d) and e)).

\begin{figure}[h]
    \centering\includegraphics[width=12cm]{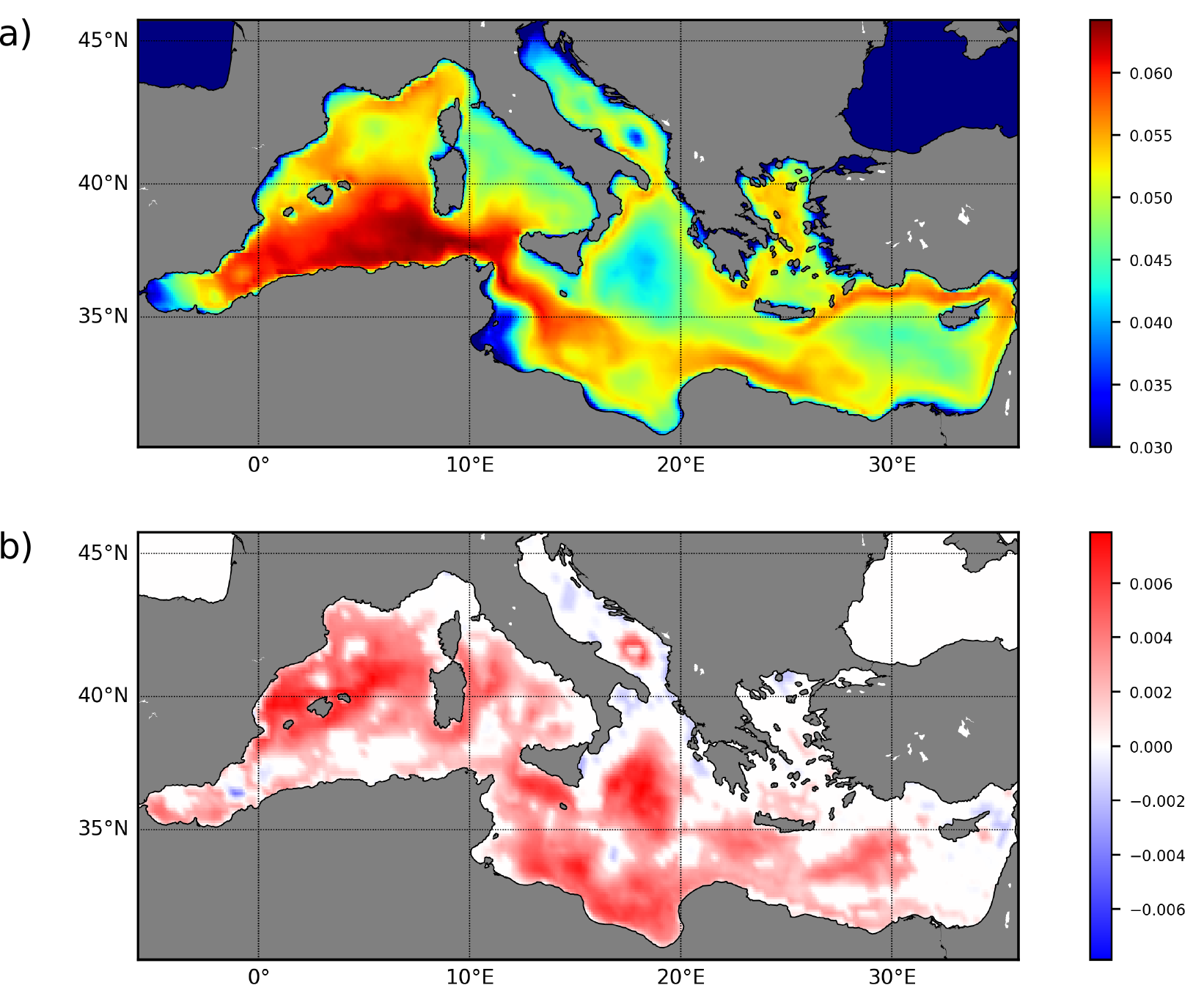}

\caption{\emph{Entropy map and significant difference}. Panel a) shows the node-scale entropy averaged over the historical period (1971-2000); panel b) shows the node-by-node statistical significant difference between future (2071-2100) and historical periods. Maps obtained with tau=60 days at weekly temporal resolution, interpolated for graphical purposes. Significance is evaluated by Welch's tests with 0.01 threshold, nodes where no significant difference were found are white.}  \label{fig:ent60d}  
\end{figure}

To assess the spatial patterns of horizontal stirring, we analyze the temporal mean of the node entropy over the historical run and the node-by-node statistical significant difference among scenario and control run (Fig. \ref{fig:ent60d}). While the increase of entropy concerns most of the Mediterranean basin, it is spatially heterogeneous and particularly pronounced in the Balearic sea, central and southern Ionian and Gulf of Sirte.  Note also that some regions show instead moderate decrease, in particular the Adriatic and Aegean Seas, the northern Ionian sea and the extreme east of the Levantine sea. Similar spatial patterns are found for $\tau=30$ and $90$ days (not shown). The same analyses performed for KE, MKE and EKE maps (Supplementary Figs. 1, 2, 3) highlight spatially-inhomogeneous but significant increases of energy across the Mediterranean basin as well.

\subsection{Relationships between entropy, kinetic and potential energy}\label{sec:res_entrenerg}

Given the similarities evidenced between kinetic energy and entropy trends, we now investigate their statistical relationship. We find that KE, MKE, EKE are strongly and non-linearly correlated with entropy (Supplementary Fig. 4), with significant Spearman coefficients ranging from 0.73 to 0.85, which are higher considering EKE and KE than with MKE. This is in accord with previous studies that found similar relations among stirring and EKE \cite{waugh2006spatial,waugh2008stirring}. While the functional relationship between KE and entropy is analogous in both runs, we also note a shift toward larger energies in the scenario run as compared to the historical one indicating that a strengthened horizontal stirring is concomitant with an increase of KE.

Moreover, comparing Fig. \ref{fig:ent60d} and Supplementary Fig. 1 we observe that the areas experiencing the largest increase of entropy tend to be also characterized by low values of kinetic energy in the historical run, nevertheless, not all the low energy areas are associated to large changes of entropy. Hence, this implies that a necessary but not sufficient condition for expecting a significant increase in stirring is to present relatively low values of kinetic energy. On the one hand, this is consistent with the non-linear relationship among KE and entropy (Supplementary Fig. 4), indicating a steep response of entropy at low values of kinetic energy that gradually saturates when the energy increases. On the other hand, large kinetic energy increases can also be suppressed by the complex bathymetry of the basin. 

To quantitatively characterize this relationship, we perform a quantile regression \cite{koenker1978regression} between KE and the relative change of entropy for each node and $\tau=30$ days (Supplementary Fig. 5). The slopes of the regression line for the 75th, 90th and 95th percentile are significantly different from zero and negative, confirming that large values of kinetic energy lower the maximum possible change in entropy. Therefore, the projected stirring increase can be ascribed to its non-linear, saturating relation with kinetic energy. Notably, this is concordant with the rising historical trend of EKE revealed from satellite observations in other regions \cite{backeberg2012impact}. They documented indeed increasing trends of EKE in the north-west Indian ocean spanning 0.005 - 0.009 $m^{2}s{-2}$ per decade. Despite drawbacks, the extrapolations of these historical trends over the next century would return EKE increases ranging from 0.05 to 0.09 $m^{2}s{-2}$, that is of the same order of the 0.05 increase of KE and a bit larger than the 0.016 rise of EKE found for a period of ten decades using the MedCORDEX simulations. 

However, the model projects a wind stress weakening across the most of the western basin and no significant increase elsewhere (Supplementary Fig. 6), indicating that wind stress can not explain by itself the KE increase. Still, the latter could be related to an increment of Available Potential Energy (APE), which can be released through baroclinic instability \cite{green1970transfer,stammer1998eddy}. Model simulations display a significant strengthening of horizontal density gradients that is reflected in the future spatial patterns of APE (Supplementary Fig. 7). Interestingly, we find a significant spatial correlation (Spearman coefficient of 0.49) among EKE and APE relative changes, suggesting that baroclinic instabilities could be the main driver of the increase of energy and thus stirring. In some regions, for example north of the Balearic Islands, where mean KE of a topographic guided flows increases significantly (Supplementary Fig. 2) mean-flow-topography interaction might also play a role in modulating EKE (Supplementary Fig. 3, 7). Note finally that other studies (also based on different models) within the MEDCORDEX initiative documented a sharpening of density gradients and a weakening of wind stress in the next century \cite{somot2006transient,thiebault2016mediterranean,soto2020evolution}.

\subsection{Global scale: hydrodynamic provinces rearrangement}\label{sec:res_provinces}

The interplay between the mean flow, mesoscale stirring and bathymetry generates complex and peculiar horizontal transport patterns in the Mediterranean that contributes to create semi-permeable boundaries separating distinct portions of its surface, called hydrodynamic provinces \cite{rossi2014hydrodynamic,ser2015flow,miron2017lagrangian}. Here, we investigate how simulated provinces would be rearranged in the future analyzing ensembles of provinces partitions generated at weekly frequency for $\tau=30, 60, 90$ days across both model runs (Sections \ref{subsec:lfnsetup}, \ref{subsec:provdef} and \ref{subsec:provmultiplepart}). An example of a partition in hydrodynamic provinces associated to a single adjacency matrix (i.e. a specific run of Infomap) is shown in Supplementary Fig. 8 along with the related coherence and mixing metrics.

Surprisingly, the mean province size (measured with the spatial mean  of province areas for partition) and internal coherence (measured by $\bar{\langle \rho \rangle}$) do not show important changes between historical and scenario runs (Table 1). This could be related to the fact that the basin-average value and spatial patterns of MKE are significantly less affected in the future than the EKE ones. This would suggest indeed that the mean flow determining the position of the most persistent province boundaries would not change importantly, thus broadly preserving the current Lagrangian geography of the Mediterranean basin.


\begin{table}
\label{tab:areaprov}
  \begin{center}    
    \scalebox{0.8}{
    \begin{tabular}{l|r|c|c|c|c|c|c} 
            
      & & \multicolumn{2}{c|}{$\tau=30$}  &\multicolumn{2}{c|}{$\tau=60$}  & \multicolumn{2}{|c}{$\tau=90$} \\ 
      \hline

	  & & \textbf{TM} & \textbf{TSD} & \textbf{TM} & \textbf{TSD} & \textbf{TM} & \textbf{TSD} \\
      \hline
	  
	  \multirow{2}{*}{\textbf{SM of provinces areas}} & Control run & $42581^{**}$ & 4314 & $71052^{**}$ & 10197 & 96494 & 16636 \\
    	    						 		  & Scenario run & $43471^{**}$ & 4445 & $72350^{**}$ & 10294 & 97021 & 16260 \\
    	    						 		  & RC & \textbf{+2.0\%} & & \textbf{+1.8\%} & & \textbf{+0.5\%} & \\
	  \hline
	  
	  \multirow{2}{*}{\textbf{SSD of provinces areas}} & Control run & $30344^{***}$ & 4599 & $64700^{***}$ & 10746 & $102238^{***}$ & 17186  \\
    	    						   & Scenario run & $54924^{***}$ & 6212 & $101072^{***}$ & 14453 & $147931^{***}$ & 22071 \\   
    	    						   & RC & \textbf{+81.0\%} & & \textbf{+56.2\%} & & \textbf{+44.7\%} & \\
	  \hline				  
	
	  \multirow{2}{*}{\textbf{Global coherence ratio} $\bar{\langle \rho \rangle}$} & Control run & $0.8008^{**}$ & 0.0928 & $0.7709^{**}$ & 0.0467 & $0.7624^{*}$ & 0.0437 \\
    	    						 		  & Scenario run & $0.7964^{**}$ & 0.0908 & $0.7660^{**}$ & 0.0437 & $0.7586^{*}$ & 0.0432 \\
    	    						 		  & RC & \textbf{-0.6\%} & & \textbf{-0.6\%} & & \textbf{-0.5\%} & \\
	  \hline
	  
	  \multirow{2}{*}{\textbf{Global mixing parameter} $\bar{\langle \mu \rangle}$} & Control run & $0.401^{***}$ & 0.01 & $0.458^{***}$ & 0.011 & $0.499^{***}$ & 0.012  \\
    	    						   & Scenario run & $0.417^{***}$ & 0.0092 & $0.472^{***}$ & 0.01 & $0.515^{***}$ & 0.01 \\   
    	    						   & RC & \textbf{+4.0\%} & & \textbf{+3.1\%} & & \textbf{+3.2\%} & \\
	  \hline				  

	  \multirow{2}{*}{\textbf{Global boundary strength} $\bar{\langle \sigma \rangle}$} & Control run & $0.2051^{***}$ & 0.0133 & $0.1691^{***}$ & 0.0163 & $0.1343^{***}$ & 0.0162  \\
    	    						   & Scenario run & $0.1892^{***}$ & 0.0121 & $0.1581^{***}$ & 0.0156 & $0.1238^{***}$ & 0.0166 \\   
    	    						    & RC & \textbf{-5.7\%} & & \textbf{-6.5\%} & & \textbf{-8.5\%} & \\
	  \hline			
	  	  
    \end{tabular}
    }
    \caption{ \emph{Province areas, coherence and mixing statistics}. Spatial means (SM) and spatial standard deviations (SSD) refer to a single partition considering all the provinces in it. Temporal means (TM) and temporal standard deviations (TSD) are instead calculated across partitions. Areas are in $km^2$. The relative change (RC) is the $(TM(future)-TM(historical))/TM(historical)$. Differences among control run and scenario run TM values are considered significant (Welch's t-test) when one or more asterisks are present in superscripts with the following convention: (***) when pvalue $< 0.001$, (**) when $0.001 <$ pvalue $< 0.01$, (*) $0.01 <$ pvalue $< 0.05$.  }
  \end{center}
\end{table}


On the contrary, as shown in Table 1, our model results project a stronger mixing in the provinces interiors (increased $\bar{\langle \mu \rangle}$ of $\sim3.4\%$), more heterogeneity of their sizes (increased spatial standard deviation of province areas in each partition of $\sim60\%$) and a decrease of the global strength of their boundaries (decreased $\bar{\langle \sigma \rangle}$ of $\sim 6.9\%$, mostly driven by length decrease). The stronger homogenization of provinces interiors due to internal mixing can be directly associated to the general increment of horizontal stirring previously highlighted; indeed, the variations documented are indeed of about 3.5\% in both cases. Such stirring increase is also likely to influence the  augmented variability of province areas in the future. While providing a clear explanation for this enhanced variability is beyond the scope of the present study, we speculate that extreme events, which are expected to become more frequent in the future, could modify largely the properties of some hydrodynamic provinces and drive this clear SSD increase. Moreover, the decrease of the global boundary strength for partition could be related directly to the increased SSD of provinces areas using heuristic geometrical relationships among perimeters and areas of 2-dimensional shapes (Section \ref{subsec:area_per}). Indeed, under reasonable approximations, for a partition in which the SM of province areas is kept constant, if the SSD increase the boundary length would decrease. In our case, the average decrease of boundary length is of 5.9\% (for $\tau=60$). Intuitively, this can be seen as if the larger provinces would become larger by eroding the boundaries with the smaller ones, making the latter even smaller and following a kind of ``rich get richer'' dynamics. This is also quantitatively supported by an 8.3\% increase of positive skewness when comparing the province areas distributions of the historical and scenario runs (for $\tau=60$).

Inspecting results from different integration time-scales, we find a marked increase of province area SM and SSD with $\tau$. This is certainly due to the fact that when $\tau$ is increasing, water parcels experiment anisotropic mixing for longer time, hence they are dispersed across larger regions \cite{ser2015flow}. Consistently, the length and strength of boundaries decrease with $\tau$. While the mixing parameter presents a significant increase with $\tau$, the coherence ratio slightly decrease. Regarding the changes between historical and scenario runs, we see that all the statistics reveal the same trends across different integration times.

Only for $\tau=60$ days, we also calculate the province statistics for different regions separately: Western Mediterranean (WM) [-6 to 10 E ; 30 to 45 N], Eastern Mediterranean (EM) [10 to 36.3 E ; 30.1 to 38.9 N], Tyrrhenian Sea (TS) [9.44 to 16.2 E ; 37.3 to 44.3 N], Adriatic Sea (AS) [12.6 to 19.7 E ; 40.3 to 40.6 N]. We find, similarly to what was reported for the whole basin, that the SM of provinces areas in each partition does not change much but that the SSD does change significantly. In particular, the latter is expected to increase of 59\% in the WM, 6\% in the EM, 22\% in the TS and 188\% in the AS, illustrating how the basin-scale signal can be amplified or alleviated in different sub-regions. Note however that low confidence lies in the large predicted increase in the AS due to its size and topography which may require even higher spatial resolution of the ocean model.

\begin{figure}[h]
    \centering\includegraphics[width=12cm]{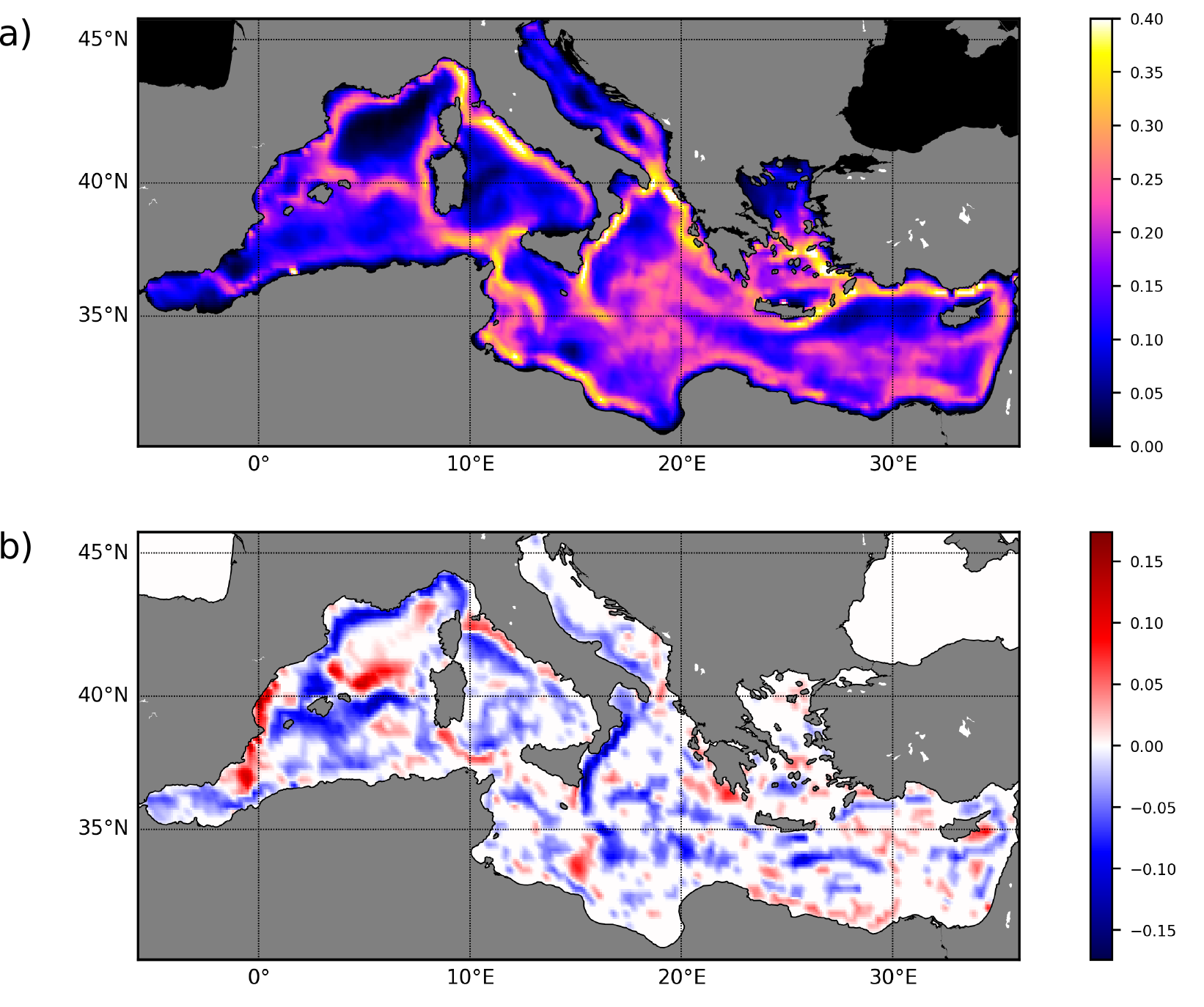}

\caption{\emph{Boundary strength map and significant difference}. Panel a) shows the boundary strength $\langle \sigma_i \rangle$ averaged over the historical run; panel b) shows the statistically significant difference among historical and scenario runs. Both fields have been interpolated for graphical purposes. Significance is evaluated by Welch's tests with threshold at 0.01, nodes with no significant difference are white colored}  \label{fig:boundstr}
  
\end{figure}

Finally, we analyze how the future rearrangement and changing properties of hydrodynamic provinces manifest themselves over the spatial dimension. In Fig.  \ref{fig:boundstr} we show the temporal mean of the node boundary strength  for $\tau=60$ days calculated at weekly frequency  and the node-by-node significant difference among both runs. In the historical run, strong boundaries are clearly identified by marked high values of boundary strength while other regions present more stochastic transport patterns denoted by smoother gradients of $\sigma$.  Such persistent boundaries are generally co-located with energetic currents and intense fronts which subdivide the ocean surface and define the main transport pathways. Their spatial locations vary with time (i.e. across partitions), especially due to intense mesoscale activity, creating these relatively large ``corridors'' of enhanced boundary occurrence \cite{rossi2014hydrodynamic}. Plotting maps of mean boundary occurrence, that is the mean probability of finding a boundary regardless its strength, we find very similar spatial patterns (confirmed by Spearman coefficients $>98\%$). Consistently with Table 1, the boundary strength map for $\tau=30$ days suggests the presence of more, smaller provinces with a denser boundary presence; on the contrary, for $\tau=90$ we can recognize larger provinces with associated sparser boundaries (Supplementary Figs. 9, 10). Moreover, note that the spatial patterns of boundary strength match qualitatively quite well eco-regionalization exercises \cite{d2009trophic,basterretxea2018patterns,ayata2018regionalisation,el2019phytoplankton}. This suggests that transport-based regionalization, like the one we provide, play a role in shaping plankton biogeography and biogeochemical regimes across the surface ocean. It is worth noting that the vertical dynamics neglected here undoubtedly affect the dispersal of micro-organisms, the distribution of dissolved chemicals as well as the long-term fate of pollutants \cite{rossi2013multi}.

When comparing both runs, in addition to the already mentioned general decreasing trend of basin boundary strength, we note that new boundaries appear in the future while others move or weaken. For instance, the boundary linked to the Liguro-Provencal current tend to decrease in strength while the one associated to the Balearic fronts is predicted to move northward and a new one is expected to appear along the Spanish coast. Considering the similitude discussed previously between the transport-based regionalization and the mean repartition of numerous active oceanic tracers, such rearrangement would also have biological and managerial implications in the future\cite{ayata2018regionalisation,hidalgo2019accounting}.

\section{Conclusions and perspectives}

Our innovative analyses applied to model projections suggest that a significant increase of entropy, as an accurate Lagrangian measure of stirring, will occur in the next century across the Mediterranean Sea. It is associated with concomitant rise of kinetic energy, mostly of its turbulent component. Such increase would saturate in the energetic areas of the basin likely due to the interplay of a non-linear relationship between entropy and kinetic energy and the effect of complex bathymetry. More energy at the Mediterranean Sea surface could be related to increases of available potential energy in its upper layer leading to more instabilities, rather than caused by changes in wind forcing. Moreover, community detection analysis allows to estimate the rearrangement of hydrodynamic provinces  and statistically estimate its synoptic consequences. It highlights a large increase of spatial standard deviation of provinces areas not followed by its associated spatial mean. Our model results project a significant increase of the global mixing parameter and decrease of the global boundary strength, the latter being most likely caused by the larger standard deviation of provinces areas i.e. largest provinces get larger, smallest provinces get smaller. Concurrently, some of the most relevant province boundaries would remain unchanged, others would move or weaken, while a few new boundaries would appear in the future. 

Our projected changes could have implications for the transport and stirring of several oceanic tracers such as nutrients, dissolved gases and, to a certain extent, drifting organisms (plankton, eggs, larvae) or floating pollutants (oil, plastic). Extensions to three-dimensional modeling frameworks will be also necessary for applications in which vertical tracer displacements can not be neglected in comparison with horizontal ones, for instance when considering longer time-scales or peculiar oceanic regions (e.g. upwelling systems). As such, our results constitute a first step toward providing tools and recommendations which, backed by more operationally-oriented multi-model approaches, may assist the design of adaptive strategies for future marine spatial planning across the world ocean.


\appendix

\acknowledgments

S. T. received funding be the European Commission (Horizon 2020, MSCA IF-2016, WACO 749699: Fine-scale Physics, Biogeochemistry and Climate Change in the West African Coastal Ocean). Modeled horizontal currents velocity fields studied here are available for download on the MedCordex repository ($https://www.medcordex.eu/search/search\_files.php$). The name of the historical dataset is: $\texttt{MED-10\_CNRM-CM5\_historical\_r8i1p1\_CNRM-RCSM4\_v1}$. The name of the RCP8.5 dataset is: $\texttt{MED-10\_CNRM-CM5\_rcp85\_r8i1p1\_CNRM-RCSM4\_v1}$. E. S-G. thanks Francesco d'Ovidio for stimulating discussions.

\end{document}